%% file: ms.tex
\documentclass[twocolumn]{aastex701}
\bibpunct{(}{)}{;}{a}{}{,}
\usepackage{amsmath}

\usepackage{hyperref}
\hypersetup{colorlinks=true, linkcolor=blue, citecolor=blue, filecolor=blue, urlcolor=blue}
\graphicspath{{./}{./plots/}}

\begin{document}

\title{Cosmological Constraints from Full-Scale Clustering and Galaxy--Galaxy Lensing with DESI DR1}
\shorttitle{Small-Scale Clustering and Lensing with DESI DR1}

\correspondingauthor{Johannes U. Lange}
\email{E-mail: jlange@american.edu}
\shortauthors{J.~U.~Lange et al.}

\input{authors.tex}

\begin{abstract}
    We present constraints on cosmic structure growth from the analysis of galaxy clustering and galaxy--galaxy lensing with galaxies from the Dark Energy Spectroscopic Instrument (DESI) Data Release 1. Our analysis includes four samples drawn from the Bright Galaxy Survey (BGS) and the Luminous Red Galaxy (LRG) target classes. Projected galaxy clustering measurements from DESI are supplemented with lensing data from the Dark Energy Survey (DES), the Kilo-Degree Survey (KiDS), and the Hyper Suprime-Cam (HSC) survey around the same targets. Our method relies on a simulation-based modeling framework using the AbacusSummit simulations and a complex halo occupation distribution model that incorporates assembly bias. We analyze scales down to $0.4 \, h^{-1} \, \mathrm{Mpc}$ for clustering and $2.5 \, h^{-1} \, \mathrm{Mpc}$ for lensing, leading to stringent constraints on $S_8 = \sigma_8 \sqrt{\Omega_\mathrm{m} / 0.3}$ and $\Omega_\mathrm{m}$ when fixing other cosmological parameters to those preferred by the CMB. We find $S_8 = 0.797_{-0.024}^{+0.023}$ and $\Omega_\mathrm{m} = 0.292 \pm 0.011$ when using lensing measurements from DES and KiDS. Similarly, for HSC, we find $S_8 = 0.791_{-0.021}^{+0.020}$ and $\Omega_\mathrm{m} = 0.300 \pm 0.009$ when assuming the best-fit photometric redshift offset suggested by the HSC collaboration. Overall, our results are in good agreement with other results in the literature while continuing to highlight the constraining power of non-linear scales.
\end{abstract}


\section{Introduction}

The advent of large galaxy surveys such as the Sloan Digital Sky Survey \citep[SDSS; ][]{Samushia2012_MNRAS_420_2102, Beutler2012_MNRAS_423_3430}, the Baryon Oscillation Spectroscopic Survey \citep[BOSS; ][]{Reid2016_MNRAS_455_1553, Ahumada2020_ApJS_249_3}, the Dark Energy Survey \citep[DES; ][]{Flaugher2015_AJ_150_150, Abbott2022_PhRvD_105_3520}, the Subaru Hyper Suprime-Cam Survey \citep[HSC; ][]{Aihara2018_PASJ_70_4, More2023_PhRvD_108_3520}, the Kilo-Degree Survey \citep[KiDS; ][]{Kuijken2015_MNRAS_454_3500, Asgari2021_AA_645_104}, and, most recently, the Dark Energy Spectroscopic Instrument \citep[DESI; ][]{Levi2013_arXiv_1308_0847, DESICollaboration2016_arXiv_1611_0036, DESICollaboration2016_arXiv_1611_0037, DESICollaboration2022_AJ_164_207, Silber2023_AJ_165_9, Guy2023_AJ_165_144, Schlafly2023_AJ_166_259, Miller2024_AJ_168_95, Poppett2024_AJ_168_245, DESICollaboration2024_AJ_167_62, DESICollaboration2024_AJ_168_58, DESICollaboration2026_AJ_171_285}, has ushered in a new era in precision measurements of the large-scale structure of the Universe. These new measurements of matter and galaxy distributions have delivered important insights into the properties dark energy and gravity as well as tests of our cosmological models \citep[see, e.g.,][]{Abbott2022_PhRvD_105_3520, Sugiyama2023_PhRvD_108_3521, Ishak2025_JCAP_09_053, Adame2025_JCAP_07_028, AbdulKarim2025_PhRvD_112_3515, Wright2025_AA_703_158, DESCollaboration2026_arXiv_2601_4559}. Alongside the advance in data quality and quantity, there has been significant effort to develop new analysis methods that can extract more cosmological information from existing data.

One promising avenue to maximize information is to extend the analysis of the galaxy and matter distribution to so-called non-linear scales, i.e., scales smaller than $\mathcal O(10 \, h^{-1} \, \mathrm{Mpc})$. Interpreting non-linear scales is challenging due to non-linear gravitational effects, the complex relationship between galaxies and dark matter halos \citep{Wechsler2018_ARAA_56_435}, and the impact of baryonic physics on the matter distribution \citep[see, e.g.,][]{vanDaalen2011_MNRAS_415_3649, Springel2018_MNRAS_475_676, vanDaalen2020_MNRAS_491_2424}. However, recent studies have shown how to address these issues through advancements in cosmological simulations and machine learning \citep[see, e.g.,][for recent applications]{Yuan2022_MNRAS_515_871, Zhai2023_ApJ_948_99, Miyatake2023_PhRvD_108_3517, Lange2023_MNRAS_520_5373, Hahn2024_NatAs_8_1457}, leading to more stringent cosmological constraints. For example, whereas a conventional analytic galaxy bias model may be applicable down to $6 \, h^{-1} \, \mathrm{Mpc}$ with point-mass marginalization \citep{Porredon2022_PhRvD_106_3530}, using a simulation-based halo occoupation distribution model, \cite{Lange2023_MNRAS_520_5373} analyzed scales down to $2.5 \, h^{-1} \, \mathrm{Mpc}$ without point-mass marginalization. Other promising directions are the analysis of higher-order clustering \citep[see, e.g.,][]{Hahn2021_JCAP_04_029, Valogiannis2024_PhRvD_109_3503, Liu2025_JCAP_05_064}, including field-level inference \citep[see, e.g.,][]{Nguyen2024_PhRvL_133_1006} and more physically-motivated priors on bias parameters \citep[see, e.g.][]{Ivanov2025_PhRvD_111_3548}.

In this work, we focus on the combined analysis of galaxy clustering and weak gravitational lensing. The clustering of galaxies is measured with data from DESI Data Release 1. We supplement these clustering measurements with lensing measurements from the latest publicly available lensing data from DES, KiDS, and HSC. Such an analysis, sometimes called a ``2$\times$2pt'' analysis, places constraints on $S_8 = \sigma_8 \sqrt{\Omega_\mathrm{m} / 0.3}$, where $\sigma_8$ is the root mean square of matter fluctuations on a scale of $8 \, h^{-1} \, \mathrm{Mpc}$ and $\Omega_\mathrm{m}$ is the matter density. The present analysis focuses on analyzing non-linear scales using advanced simulation-based modeling. It is complementary to another study by \cite{Porredon2026_OJAp_9_1342} that analyzes the matter and galaxy distribution on larger scales and also includes cosmic shear data in a so-called ``3$\times$2pt'' analysis.

Our analysis provides new insights into the so-called $S_8$-tension, a possible discrepancy between inferred values of $S_8$ between studies of the Cosmic Microwave Background (CMB) and low-redshift probes such as gravitational lensing \citep{Abdalla2022_JHEAp_34_49}. While such a tension may also exist on larger scales \citep{Abbott2022_PhRvD_105_3520, Dalal2023_PhRvD_108_3519, Wright2025_AA_703_158}, non-linear scales are particularly sensitive to it. Early work by \cite{Cacciato2013_MNRAS_430_767}, who studied lensing and clustering down to small scales, found values for $S_8$ significantly below than that preferred by the \cite{PlanckCollaboration2020_AA_641_6} CMB analysis, $S_8 = 0.825 \pm 0.011$. Whereas the analysis by \cite{Cacciato2013_MNRAS_430_767} was based on an analytic halo model, many recent studies employ simulation-based modeling to predict galaxy clustering and lensing. One such example is the study by \cite{Leauthaud2017_MNRAS_467_3024} that showed that the lensing amplitude at fixed clustering is significantly overpredicted when one assumes cosmological parameters preferred by CMB studies. However, this discrepancy may also be related to galaxy physics such as baryonic feedback and galaxy asssembly bias \citep{Leauthaud2017_MNRAS_467_3024, Lange2019_MNRAS_488_5771, Amon2022_MNRAS_516_5355, ChavesMontero2023_MNRAS_521_937}. Subsequent studies mitigated the impact of unknown galaxy physics through scale cuts but generally continued to find low values for $S_8$. For example, \cite{Miyatake2023_PhRvD_108_3517}, analyzing HSC data together with clustering measurements from BOSS found $S_8 = 0.763^{+0.040}_{-0.036}$. Similarly, the analysis of BOSS clustering and DES and KiDS lensing by \cite{Lange2023_MNRAS_520_5373} yielded $S_8 = 0.792 \pm 0.022$ \citep[also see][]{Wibking2020_MNRAS_492_2872, Dvornik2023_AA_675_189, Amon2023_MNRAS_518_477}. In this work, we build upon these studies using new measurements from DESI, the successor to BOSS.

We begin by outlining our data and measurements in section~\ref{sec:measurements}. Next, we describe our analysis method, including tests on simulated mock catalogs, in section~\ref{sec:methods}. Our main results are presented in section~\ref{sec:results} before discussing and summarizing our findings in section~\ref{sec:conclusion}.

\section{Measurements}
\label{sec:measurements}

Our analysis relies on galaxy clustering measurements from DESI Data Release 1, supplemented with gravitational lensing measurements from DES Year 3 \citep[DES Y3; ][]{Gatti2021_MNRAS_504_4312}, the KiDS Data Release 4 \citep[KiDS-1000; ][]{Giblin2021_AA_645_105}, and HSC Year 3 \citep[HSC Y3; ][]{Li2022_PASJ_74_421} data. The measurements are described in detail in \cite{Heydenreich2025_arXiv_2506_1677} and we summarize the most salient points here. All measurements assume $\Omega_\mathrm{m} = 0.3075$ \citep{PlanckCollaboration2016_AA_594_24} to convert redshifts and angles into physical coordinates.

\subsection{DESI Samples}

The different galaxy samples were investigated in \cite{Yuan2024_MNRAS_533_589} and balance increased number density with the need for negligible redshift evolution within each sample. DESI targets different classes of extragalactic objects. Those at lower redshift, $z < 1$, i.e., those for which we can get precise galaxy--galaxy lensing measurements, are the Bright Galaxy Survey (BGS) and the Luminous Red Galaxy (LRG) sample.

The BGS Bright targets include all objects with an apparent $r$-band magnitude brighter than $19.5$ \citep{Hahn2023_AJ_165_253} and covers an area of around $7,500 \, \mathrm{deg}^2$ \citep{Adame2025_JCAP_07_017}. Due to being a flux-limited survey, the BGS includes many low-redshift targets with a strongly redshift-dependent number density. As a result, we need to apply additional cuts on the absolute k-corrected and evolution-corrected magnitudes $M_\mathrm{r}$ \citep{Moustakas2023_ascl_soft_8005, DESICollaboration2024_AJ_168_58}. We have three BGS samples, BGS1, BGS2, and BGS3. Their redshift ranges are defined by the boundaries $[0.1, 0.2, 0.3, 0.4]$ and the absolute luminosity thresholds are $M_\mathrm{r} < -19.5$, $-20.5$, and $-21$, respectively. However, we do not use BGS1 in our analysis since our simulations do not have sufficient mass resolution to model this low-luminosity sample. In particular, the number density of BGS1 significantly exceeds the number of resolved halos in the simulation. Since the number density of central galaxies cannot exceed that of resolved halos, this forces a very high satellite fraction in the fit and an overall unacceptable $\chi^2$. A set of higher-resolution simulations would be needed to effectively model this sample.

The LRG sample has a slightly more complicated selection based on the $g$, $r$, $z$, and $W1$ magnitudes \citep{Zhou2023_AJ_165_58} and covers an area of around $5,700 \, \mathrm{deg}^2$. Fortunately, the LRG selection does not result in a strongly redshift-dependent number density. Thus, we define three LRG samples, LRG1, LRG2, and LRG3, by the redshift boundaries $[0.4, 0.6, 0.8, 1.1]$. However, for our main analysis, we do not use LRG3 since these measurements are potentially strongly contaminated by intrinsic alignments, lens magnification and other lensing systematics \citep{Heydenreich2025_arXiv_2506_1677}.

For all samples, we use the galaxy number density $n_\mathrm{gal}$ as a constraint. The number density, like the clustering measurements described below, are corrected for incompleteness in the DESI fiber assignment using weights. Uncertainties are derived from jackknife resampling. The number densities are $(3.71 \pm 0.03) \times 10^{-3}$, $(1.38 \pm 0.01) \times 10^{-3}$, $(5.70 \pm 0.03) \times 10^{-4}$, and $(5.69 \pm 0.03) \times 10^{-4} \, h^3 \, \mathrm{Mpc}^{-3}$ for BGS2, BGS3, LRG1, and LRG2, respectively.

\subsection{Galaxy Clustering}

Galaxy clustering is measured via the projected two-point correlation function $w_\mathrm{p}$,
\begin{equation}
    w_\mathrm{p}(r_\mathrm{p}) = 2 \int\limits_0^{\pi_\mathrm{max}} \xi(r_\mathrm{p}, r_\pi) \, \mathrm{d} r_\mathrm{p} \, .
\end{equation}
In the above equation, $\xi(r_\mathrm{p}, r_\pi)$ denotes the redshift-space two-point correlation function as a function of the perpendicular, $r_\mathrm{p}$, and line-of-sight direction, $r_\pi$. By integrating along the line-of-sight direction out to $\pi_\mathrm{max} = 80 \, h^{-1} \, \mathrm{Mpc}$ we reduce sensitivity to redshift-space distortions. The clustering measurements are corrected for fiber incompleteness via pairwise-inverse-probability (PIP) weights and angular upweighting \citep{Bianchi2018_MNRAS_481_2338, Bianchi2025_JCAP_04_074}.

We measure $w_\mathrm{p}$ in $14$ logarithmic, comoving $r_\mathrm{p}$ bins going from $10^{-1} \, h^{-1} \, \mathrm{Mpc}$ to $10^{1.8} \, h^{-1} \, \mathrm{Mpc}$ but only use data above $0.4 \, h^{-1} \, \mathrm{Mpc}$ for our analysis. Uncertainties are estimated from jackknife resampling with $128$ equal-area patches on the sky. Finally, for the inverse of the covariance matrix, we use the Hartlap correction factor \citep{Hartlap2007_AA_464_399}.

\subsection{Gravitational Lensing}

We also measure the galaxy--galaxy lensing signal around DESI samples to probe the mass distribution around them. Galaxy shape measurements come from the publicly available DES, KiDS, and HSC catalogs. The DES weak lensing catalogs are based on the 3-year data set, utilize four tomographic bins based on $riz$ colors, have an effective number density of $6 \, \mathrm{arcmin}^{-2}$, and span an area of around $4,100 \, \mathrm{deg}^2$. KiDS uses a more extensive nine-band photometry to define its five tomographic bins and has a similar number density over a smaller $800 \, \mathrm{deg}^2$ area. Given its area, it is called the KiDS-1000 data set. Finally, the HSC year-3 data is significantly deeper with a number density of $15 \, \mathrm{arcmin}^{-2}$, four tomographic bins derived from five optical bands, but a much smaller $433 \, \mathrm{deg}^2$ area.

The galaxy--galaxy lensing amplitude is a measure of the so-called excess surface density $\Delta\Sigma$ around galaxies,
\begin{equation}
    \Delta\Sigma(r_\mathrm{p}) = \frac{2}{r_\mathrm{p}^2} \int\limits_0^{r_\mathrm{p}} \hat{r}_\mathrm{p} \, \Sigma(\hat{r}_\mathrm{p}) \, \mathrm{d} \hat{r}_\mathrm{p} - \Sigma(r_\mathrm{p}) \, ,
\end{equation}
where $\Sigma(r_\mathrm{p})$ is the projected mass density at a certain distance $r_\mathrm{p}$ perpendicular to the line of sight. In practice, $\Delta\Sigma$ is inferred from the mean tangential shear $\gamma_\mathrm{t}$,
\begin{equation}
    \gamma_\mathrm{t}(r_\mathrm{p}) = \frac{\Delta\Sigma(r_\mathrm{p})}{\Sigma_\mathrm{crit}(z_\mathrm{l}, z_\mathrm{s})}
\end{equation}
of ``source'' galaxies in DES Y3, KiDS-1000, and HSC Y3 around DESI ``lens'' galaxies. In the above equation, $\Sigma_\mathrm{crit}$ is the critical surface density and depends on both the lens, $z_\mathrm{l}$, and the source redshift, $z_\mathrm{s}$. We refer the reader to \cite{Heydenreich2025_arXiv_2506_1677} for a detailed discussion of how these measurements are performed. These measurements include corrections for fiber incompleteness using individual inverse probably weights which have been shown to be accurate to better than $\lesssim 1\%$ in simulations \citep{Lange2024_OJAp_7_57}. Additionally, \cite{Heydenreich2025_arXiv_2506_1677} found no residual correlation between measured lensing amplitudes and the number of overlapping tiles for each galaxy. This indicates that correction for fiber incompleteness is robust. We use the lensing measurements employing the conservative lens-source cuts, as outlined in Table 1 of \cite{Heydenreich2025_arXiv_2506_1677}. Employing such cuts reduces systematic uncertainties related to, among others, intrinsic alignments and boost factors \citep{Lange2024_OJAp_7_57}. This implies that we do not use DES and KiDS lensing measurements for LRG2. We correct for lens magnification \citep{Unruh2020_AA_638_96} assuming the best-fit \cite{PlanckCollaboration2020_AA_641_6} cosmological parameters. Since lens magnification is a small, $\lesssim 10\%$ effect, we neglect its cosmological dependence. We also do not explicitly correct for intrinsic alignment contamination since the effect is expected to be small given that it is only relevant if lenses and sources are at the same redshift. Using the data from \cite{Lange2024_OJAp_7_57}, we estimate that for our lens-source cuts and a minimum scale of $r_\mathrm{p} = 2.5 \, h^{-1} \, \mathrm{Mpc}$, the impact is well below $\Delta \chi^2 < 1$ with most of the significance occuring for LRG1 lenses and source from DES and KiDS. That said, there is substantial uncertainty on the magnitude of intrinsic alignments and future studies using higher-precision lensing measurements should investigate potential corrections for intrinsic alignments.

We measure $\Delta\Sigma$ in $13$ logarithmic, comoving bins going from $10^{-1} \, h^{-1} \, \mathrm{Mpc}$ to $10^{1.6} \, h^{-1} \, \mathrm{Mpc}$. To avoid sensitivity to baryonic effects and subhalo lensing \citep[see, e.g.,][]{Zu2015_MNRAS_454_1161, Lange2019_MNRAS_488_5771, Amodeo2021_PhRvD_103_3514}, we only use scales larger than $2.5 \, h^{-1} \, \mathrm{Mpc}$ for our analysis. While baryonic effects will also impact the distribution of satellite galaxies and thereby clustering, we marginalize over this in our analysis. This is why we employ a lower minimum scale for clustering than lensing. We estimate uncertainties using the same jackknife procedure as for clustering. We ignore possible correlations between clustering and lensing amplitudes. Those are expected to be small \citep{Taylor2022_PhRvD_106_3536}, especially given that clustering is measured over the full DESI footprint, $7,500 \, \mathrm{deg}^2$ and $5,700 \, \mathrm{deg}^2$ for BGS and LRG, respectively, whereas the lensing amplitude measurements are limited to $\lesssim 1,000 \, \mathrm{deg}^2$ overlap areas with the lensing surveys.

For our cosmological analysis, we combine lensing measurements made for the same DESI lenses but different source samples, i.e., different lensing surveys. Specifically, if we have $n$ lensing estimates $\mu_i$ with associated covariances $\Sigma_i$, the combined covariance is
\begin{equation}
    \Sigma_\mathrm{tot} = \left( \sum\limits_{i=1}^n \Sigma_i^{-1} \right)^{-1}
\end{equation}
and the combined estimate becomes
\begin{equation}
    \mu_\mathrm{tot} = \Sigma_\mathrm{tot} \sum\limits_{i=1}^n \Sigma_i^{-1} \mu_i \, .
\end{equation}
We derive two sets of combined lensing measurements for each of the DESI samples. One set includes all lensing measurements from DES and KiDS and the other one the HSC measurements. Within the DESI footprint, DES and KiDS do not overlap spatially, allowing us to treat these measurements as independent. On the other hand, the HSC footprint in DESI overlaps with both DES and KiDS. Thus, the combined DES and KiDS lensing measurements have a non-zero covariance with the HSC measurements. In the following, we will refer to the combined DES and KiDS lensing measurements as DES/KiDS. The DES/KiDS footprints have a roughly $\sim 1,000 \, \mathrm{deg^2}$ overlap with DESI Y1 while HSC Y3 has a smaller $\sim 400 \, \mathrm{deg^2}$ overlap but higher source density.

Finally, the HSC Y3 lensing measurements are significantly affected by uncertainties in the photometric redshift calibration \citep{Rau2023_MNRAS_524_5109}. In their 3$\times$2pt analysis, \cite{Li2023_PhRvD_108_3518} used a self-calibration to show that applying an additional offset of $+0.115$ and $+0.192$ to the mean redshift of the third and forth tomographic bin, respectively, significantly improves the fit. A similar result was found by \cite{Heydenreich2025_arXiv_2506_1677}. Thus, by default, we assume these offsets to make our HSC Y3 lensing measurements. Later, we also discuss how our results would change if no offsets were applied to the fiducial HSC Y3 calibration in \cite{Rau2023_MNRAS_524_5109}.

\section{Methods}
\label{sec:methods}

For the most part, our analysis method closely follows previous work by \cite{Lange2022_MNRAS_509_1779, Lange2023_MNRAS_520_5373} which we summarize here briefly.

\subsection{Simulations}

We start with a set of cosmological $N$-body simulations spanning a range of cosmologies, i.e., cosmological parameters. In previous work, the Aemulus Alpha simulations \citep{DeRose2019_ApJ_875_69} were used whereas we employ the AbacusSummit suite \citep{Garrison2021_MNRAS_508_575, Maksimova2021_MNRAS_508_4017} in this work primarily due to its larger volume and higher resolution. From AbacusSummit, we select all cosmologies that vary $\Omega_\mathrm{m}$ and $\sigma_8$ with respect to the base \cite{PlanckCollaboration2020_AA_641_6} CMB cosmology while keeping neutrino and dark energy properties as well as the shape of primordial power spectrum, i.e., the spectral index $n_s$ and the running of the spectral index, fixed. More explicitly, we use the cosmologies $0$, $1$, $4$, $100$, $101$, $102$, $103$, $112$, $113$, $116$, $117$, $118$, $125$, $126$, $130$, $131$, $133$, and $134$. Furthermore, we add the cosmologies $137$, $140$, and $144$. While these do vary the spectral index $n_s$, they add important coverage in the $S_8$-$\Omega_\mathrm{m}$ plane. The $S_8$-range probed by these simulations is $0.73$ to $0.93$ whereas $\Omega_\mathrm{m}$ goes from $0.255$ to $0.388$.

We note that all cosmological parameters besides $\Omega_\mathrm{m}$ and $S_8$ are assumed fixed in our analysis. The spectral index, with the exception of simulations $137$, $140$, and $144$ is fixed at $n_s = 0.9649$ \citep{PlanckCollaboration2020_AA_641_6}. Similarly, there is no running of the spectral index, the sum of the neutrino masses is $0.06 \, \mathrm{eV}$, $N_\mathrm{eff} = 3.0328$, and dark energy is assumed to be constant, $w_0 = -1$ and $w_a = 0$. Note that, conversly, the Hubble parameter $h = H_0 / (100 \, \mathrm{km} / \mathrm{s} / \mathrm{Mpc})$ is varied in these simulations, primarily because the CMB accoustic scale $\theta_*$ is kept fixed when varying $\Omega_\mathrm{m}$. In principle, this implies that $H_0$ and $\Omega_\mathrm{m}$ are varied together. However, since gravitational lensing is much more sensitive to $\Omega_\mathrm{m}$ than $H_0$, we interpret all changes in the goodness-of-fit as constraints on $\Omega_\mathrm{m}$ and $S_8$. Ultimately, the main focus of this work is testing the consistency between the CMB and low-redshifts measurements of cosmic structure, i.e., $S_8$ and $\Omega_\mathrm{m}$. Thus, for this comparison, we are not interested in varying other cosmological parameters beyond those ranges permitted by the CMB. Since constraints on $S_8$ are not strongly degenerate with other parameters in the ranges allowed by the CMB \citep[see, e.g.,][]{Gao2026_ApJ_998_268}, keeping them fixed our analysis seems unlikely to significantly affect this consistency assessment but should be kept in mind as a caveat.

For each cosmology, we select the base simulation with $6192^3$ particles and a volume of $(2 \mathrm{h}^{-1} \, \mathrm{Gpc})^3$, resulting in a particle resolution of $2.11 \times 10^9 \, (\Omega_\mathrm{m} / 0.314) \, \mathrm{h}^{-1} M_\odot$. For each simulation, we use catalogs of all halos \citep{Hadzhiyska2022_MNRAS_509_501} with at least $300$ particles and a random $0.025 \%$ of all particles to sample the lensing field. These catalogs are extracted at redshifts $0.2$, $0.3$, $0.4$, $0.5$, and $0.8$\footnote{Cosmology $103$ does not have data for redshift $0.8$. Therefore, we do not use that cosmology for fits involving the high-redshift LRG2 sample.}.

\subsection{Galaxy Model}

\begin{table}
    \centering
    \begin{tabular}{l|l|l}
    Parameter & Description & Range \\
    \hline
    $\log M_{\rm min}$ & Mass at which $\langle N_{\rm cen} \rangle = 0.5$ & $[11, 15]$ \\
    $\sigma_{\log M}$ & Low-mass transition of $\langle N_{\rm cen} \rangle$ & $[0.1, 1.0]$ \\
    $f_\Gamma$ & LRG incompleteness of $\langle N_{\rm cen} \rangle$ & $[0.5, 1.0]$ \\
    $\log M_0$ & Mass below which $\langle N_{\rm sat} \rangle = 0$ & $[11, 15]$ \\
    $\log M_1$ & Mass at which $\langle N_{\rm sat} \rangle \approx 1$ & $[11, 15]$ \\
    $\alpha$ & Power-law index of $\langle N_{\rm sat} \rangle$ & $[0.5, 2.0]$ \\
    $A_{\rm cen}$ & Central assembly bias & $[-1.0, 1.0]$ \\
    $A_{\rm sat}$ & Satellite assembly bias & $[-1.0, 1.0]$ \\
    $\log \eta$ & Satellite spatial bias & $[-0.5, 0.5]$ \\
    $\alpha_{\rm s}$ & Satellite velocity bias & $[0.8, 1.2]$ \\
    \end{tabular}
    \caption{Description and ranges of galaxy parameters varied in our analysis.}
    \label{tab:galaxy_parameters}
\end{table}

Halos in the AbacusSummit simulations are populated with galaxies via a Halo Occupation Distribution \citep[HOD; ][]{Berlind2002_ApJ_575_587, Bullock2002_MNRAS_329_246, Zheng2007_ApJ_667_760} model. The number of centrals, $N_{\rm cen}$, per dark matter halo as a function of its halo mass $M$ is given by
\begin{equation}
    \langle N_{\rm cen} | M \rangle = \frac{f_\Gamma}{2} \left( 1 + \mathrm{erf} \left[ \frac{\log M - \log M_{\rm min}}{\sigma_{\log M}} \right] \right),
\end{equation}
whereas the number of satellites, $N_{\rm sat}$, obeys
\begin{equation}
    \langle N_{\rm sat} | M \rangle = \left( \frac{M - M_0}{M_1} \right)^\alpha \, .
\end{equation}
In the above equations, $f_\Gamma$, $\log M_{\rm min}$, $\sigma_{\log M}$, $M_0$, $M_1$, and $\alpha$ are free parameters that we marginalize over in our analysis. The parameter $f_\Gamma$ accounts for potential color incompleteness at high stellar masses for LRGs and is fixed to unity for the BGS samples.

The above parametrization describes a simple mass-only HOD and is supplemented with the decorated HOD framework \citep{Hearin2016_MNRAS_460_2552} to allow for galaxy assembly bias based on halo concentration. This gives two more free parameters, $A_\mathrm{cen}$ and $A_\mathrm{sat}$, controlling the amount of assembly bias for centrals and satellites, respectively. Specifically, the number of centrals is modified based on the halo concentration $c$ according to
\begin{equation}
    \begin{split}
        \langle N_{\rm cen} | M, c \rangle &=\\
        \langle N_{\rm cen} | M \rangle &\pm A_{\rm cen} \, \mathrm{min} (\langle N_{\rm cen} | M \rangle, 1 - \langle N_{\rm cen} | M \rangle) \, .
    \end{split}
\end{equation}
In the above equation, the second term is added if the halo concentration is above the median concentration for halos of that mass and subtracted otherwise. Similarly, the satellite number becomes
\begin{equation}
        \langle N_{\rm sat} | M, c \rangle = \langle N_{\rm sat} | M \rangle \pm A_{\rm sat} \, \langle N_{\rm sat} | M \rangle \, .
\end{equation}
Both $A_{\rm cen}$ and $A_{\rm sat}$ are limited to the range $[-1, +1]$ to ensure the expected number of galaxies is always non-negative.

Finally, satellite positions and velocities are drawn from an analytic Navarro–Frenk–White \citep[NFW; ][]{Navarro1997_ApJ_490_493} profile with the second moment of the velocity distribution calculated from the time-independent Jeans equation. We allow satellites to have a biased spatial and velocity distribution with respect to the dark matter within each halo \citep[see][for details]{Lange2023_MNRAS_520_5373}. The spatial bias is defined by $\eta = c_\mathrm{sat} / c$, where $c_\mathrm{sat}$ is the concentration parameter of the satellite distribution within a halo. Similarly, $\alpha_\mathrm{s}$ controls the velocity bias and represents an additional factor to multiply satellite velocities with.

We list all galaxy parameters together with a short description and the range of values we explore in our analysis in Table~\ref{tab:galaxy_parameters}. The ranges explored are very broad and, in many cases, cover all values that meaningfully affect the predictions while being consistent with observations. Similarly, the limits for the assembly bias parameters are the maximum ranges allowed by the model \citep{Hearin2016_MNRAS_460_2552}. While the ranges for $\log \eta$ and $\alpha_s$ could, in principle, be expanded, these parameters are virtually unconstrained by the data.

The above model allows us to place simulated galaxies into dark matter-only simulations. We then use the distant-observer approximation to measure galaxy clustering and lensing. For clustering, we take redshift-space distortions into account. For lensing, we additionally make use of down-sampled particle catalogs to estimate the excess surface density $\Delta\Sigma$, using the algorithm outlined in \cite{Lange2019_MNRAS_488_5771}. All mock measurements use the same binning as the real data and the clustering predictions include the Alcock-Paczyński effect \citep{Alcock1979_Natur_281_358}, i.e., the assumed cosmology when converting redshifts and angles to physical coordinates. For lensing, the effect is expected to be small \citep{More2013_ApJ_777_26} and is ignored.

We accelerate the prediction of galaxy clustering and lensing via {\sc TabCorr} version $1.1.2$. This software utilizes the tabulation method described in \cite{Zheng2016_MNRAS_458_4015, Lange2019_MNRAS_490_1870}. In essence, we first calculate the clustering and lensing of central and satellite galaxies within halo mass and concentration bins. Afterward, the correlation functions are convolved with the HOD to make predictions for the galaxy correlation functions. Finally, we use cubic spline interpolation to make predictions for arbitrary phase-space parameters, i.e., $\log \eta$ and $\alpha_s$, and redshifts. We checked that interpolation errors are small compared to observational uncertainties. Finally, we use the `efficient` configuration in {\sc TabCorr} to precompute the halo correlation functions.

Finally, we note two shortcomings in our galaxy model. Most importantly, we have implicitly neglected the impact of baryonic physics on the galaxy--galaxy lensing amplitude \citep[see, e.g.,][]{Leauthaud2017_MNRAS_467_3024, Lange2019_MNRAS_488_5771, Amodeo2021_PhRvD_103_3514, BeltzMohrmann2021_ApJ_921_112, Amon2023_MNRAS_518_477, Bigwood2024_MNRAS_534_655, Sunseri2026_MNRAS_546_g031}. Typically, neglecting baryonic processes will lead to an overprediction of the lensing amplitude on small scales. The scale cuts we impose are intended to mitigate this issue. We tested this by implementing the baryonic suppression of the lensing amplitude found by \cite{Lange2019_MNRAS_488_5771} for galaxies above a mass of $10^{11} \, M_\odot$ in the original Illustris simulation \citep{Vogelsberger2014_MNRAS_444_1518}. Notably, the Illustris simulation already predicts a strong suppression due to baryonic feedback compared to other simulations \citep{Amon2022_MNRAS_516_5355}. Nonetheless, we find that the baryonic suppression changes the final DES/KIDS constraints on $S_8$ and $\Omega_\mathrm{m}$ by less than $0.3 \sigma$.

In addition to baryonic feedback, by drawing satellite positions from a smooth analytic density profile, we implicitly neglect the impact of subhalos \citep[see, e.g.,][]{Zu2015_MNRAS_454_1161, ChavesMontero2023_MNRAS_521_937} on the lensing amplitude. Unlike baryonic feedback, this simplification will lead to an underprediction of the lensing signal on small scales. However, as for baryonic feedback, the scale cut should substantially mitigate this issue. Finally, we implement assembly bias based on halo concentration as the secondary halo property. However, recent works advocate for using other secondary halo properties such a over-density to model galaxy assembly bias \citep{Xu2021_MNRAS_502_3242, Hadzhiyska2021_MNRAS_508_698}. Further tests are needed to determine whether using a different secondary halo property has a strong impact on our results.

\subsection{Cosmological Inference}

Starting with a set of observational data, $D_\mathrm{obs} = \{ n_\mathrm{gal}, w_\mathrm{p}, \Delta\Sigma \}$, our goal is to infer constraints on $\Omega_\mathrm{m}$ and $S_8$. The first step is to determine the best-fit, i.e., maximum-likelihood, galaxy model $D_\mathrm{mod}$ for each AbacusSummit simulation with respect to the observed data. The likelihood is defined as
\begin{equation}
    \begin{split}
        \log \mathcal{L} &= - \frac{1}{2} \chi^2 \\
        &=- \frac{1}{2} \left( D_\mathrm{obs} - D_\mathrm{mod} \right)^\mathrm{T} \Sigma^{-1} \left( D_\mathrm{obs} - D_\mathrm{mod} \right) \, ,
    \end{split}
\end{equation}
where $\Sigma$ is the covariance matrix of the observational data. We determine the maximum-likelihood galaxy model and posterior for each simulation using {\sc nautilus} \citep{Lange2023_MNRAS_525_3181} with $3,000$ live points and a stopping criterion of $f_\mathrm{live} < 10^{-10}$. To determine cosmology constraints, in previous works \citep[see, e.g., ][]{Lange2023_MNRAS_520_5373}, we ``emulated'' the maximum likelihood $\log \mathcal{L}_\mathrm{max}$ as a function of cosmology via skew-normal distributions. In this work, we opt for a different approach where we instead emulate the maximum-likelihood model predictions as a function of cosmology. We choose this method for its computational efficiency and to have a different method to test against. Additionally, we find this method to be more precise, resulting in less uncertainty from the limited number of simulations. The observational $S_8$ constraints in this work do not shift significantly with the skew-normal approach but have around $30\%$ larger uncertainties.

For each simulation representing different values for $\Omega_\mathrm{m}$ and $S_8$, we have the best-fit prediction $D_\mathrm{m}$.\footnote{Cosmologies $137$, $140$, and $144$ vary $n_s$ compared to all other cosmologies. We correct for this by calculating $\partial D_\mathrm{m} / \partial n_s$ using cosmologies $104$ and $105$ and subtracting the expected contribution of the $n_s$ variations assuming linearity. This correction was added post-unmasking when we found that the best-fit models of $137$, $140$, and $144$ were not well predicted from the other cosmologies. Our results change by at most $0.2 \sigma$ if we do not apply these corrections or do not use $137$, $140$, and $144$ at all.} First, we normalize these data vectors via $\tilde D_\mathrm{mod} = D_\mathrm{mod} / \sqrt{\mathrm{diag}(\Sigma)}$ and reduce the dimensionality via a Principal Component Analysis (PCA). We only keep those $N_\mathrm{PCA}$ eigenvectors $u_i$ that explain $99.9\%$ of the variance across cosmology, typically the first $3$ to $6$ eigenvectors. Finally, we fit the cosmology dependence of each PCA component $x_i$ via a simple linear model,
\begin{equation}
    \hat x_i = a_i + b_i S_8 + c_i \Omega_\mathrm{m} + d_i S_8^2 \, .
    \label{eq:linear_model}
\end{equation}
We chose this specific linear model because it closely corresponds to a skew-normal likelihood, with a skew in the $S_8$-dimension. This allows us to estimate the best-fit model $D_\mathrm{mod}$, and thereby best-fit likelihood, for arbitrary values of $\Omega_\mathrm{m}$ and $S_8$. To account for uncertainties in this interpolation, we measure the leave-one-out scatter $\sigma^2_i = \mathrm{Var}(x_i - \hat x_i)$ between the linear model and the actual values of the PCA components and add this to the covariance matrix,
\begin{equation}
    \Sigma \rightarrow \Sigma + \sum\limits_{i=1}^{N_\mathrm{PCA}} \sigma_i^2 \left( u_i \sqrt{\mathrm{diag}(\Sigma)} \right)^\mathrm{T} \left( u_i \sqrt{\mathrm{diag}(\Sigma)} \right) \, .
\end{equation}
For the DESI observational data, the emulation error is significantly smaller than the observational error for all samples and measurements, giving confidence that our results are not strongly affected by emulation uncertainties. For example, for the DESI observational data, $\sigma_\mathrm{emu} \lesssim 0.4 \, \sigma_\mathrm{obs}$ for all samples and measurements. As an additional test, we sample the best-fit coefficients of the linear fit according to its covariance estimated from the linear regression. This exercise indicates that the uncertainty of the maximum-likelihood estimate due to the interpolation is a factor of $\sim 5 - 10$ smaller than observational uncertainties. As a final test, we find that our results do not sensitively depend on the linear model or amount of variance considered. We tested a more flexible model in equation \eqref{eq:linear_model} by adding an $S_8 \times \Omega_{\mathrm m}$ and an $S_8^3$ term but do not find significant differences for our observational contraints. All this suggests that our results are not strongly affected by uncertainties in the interpolation over cosmology.

\subsection{Verification}

Before applying our analysis method to (masked) DESI data, we perform tests on simulated mock data to check for potential biases in our analysis method. The mock measurement are based on the DESI emulator mock challenge (Beltz-Mohrmann et al. in prep.). Mock galaxy catalogs are generated via the GalSampler method \citep{Hearin2020_MNRAS_495_5040} by transferring the galaxy-halo connection model from the diffksy model \citep{Alarcon2023_MNRAS_518_562, Hearin2023_MNRAS_521_1741} run on SMDPL simulation \citep{Prada2012_MNRAS_423_3018} to the UNIT simulations \citep{Chuang2019_MNRAS_487_48}.

As part of the DESI emulator mock challenge, LRG-like mocks were created at redshifts $0.5$ and $0.8$. Thus, to most closely mimic our analysis, we choose the $0.5$ mock. However, that means that, unlike for the analysis of DESI data, we only use a single galaxy sample. From the available galaxy models, we choose the one where the GalSampler method use both halo mass and concentration. Furthermore, the intra-halo phase-space coordinates for galaxies use the host-centric model. We refer the reader to Beltz-Mohrmann et al. (in prep.) for more details about these simulations.

The mock catalogs are tuned to reproduce the properties of the BGS and LRG samples in the DESI Early Data Release \citep{DESICollaboration2024_AJ_168_58}. However, they do not include projections onto the sky or the actual DESI footprint. Instead, we use the distant observer approximation, averaging the results from the three simulation axes chosen as the line of sight. The covariance matrix used for inference is estimated using the jackknife method with $125$ equal-volume parts each of the two UNIT simulation mocks. We note that uncertainties in these mock measurements, especially $\Delta\Sigma$, are much smaller than observational ones.

\begin{figure}
    \centering
    \includegraphics{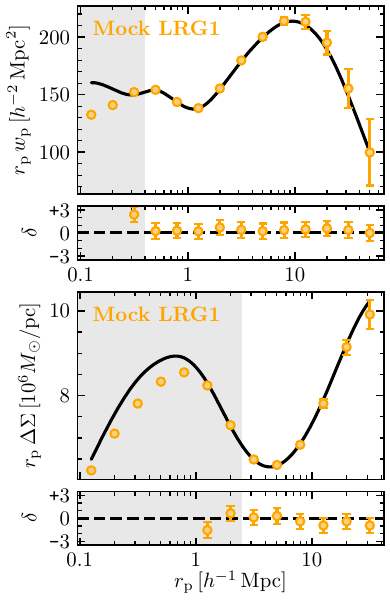}
    \caption{Best-fitting clustering and lensing predictions for the mock LRG1 measurements. The lower smaller panels indicate the difference between observations and model in units of the observational uncertainty. Gray regions indicate measurements not used in the fit.}
    \label{fig:mock_best_fit}
\end{figure}

\begin{figure}
    \centering
    \includegraphics{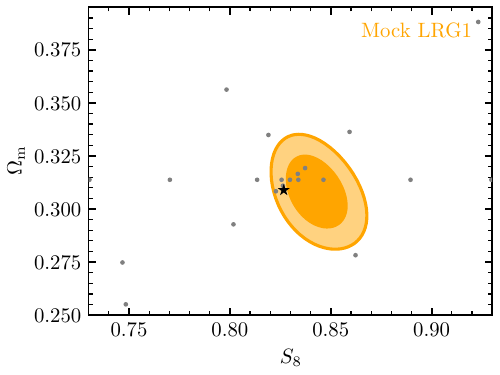}
    \caption{Mock constraints on $S_8$ and $\Omega_\mathrm{m}$. The different shaded regions indicate parameter spaces where $0 \leq \Delta \chi^2 < 2.28$ and $2.28 \leq \Delta \chi^2 < 5.99$ with respect to the best-fit cosmology, corresponding to the $68\%$ and $95\%$ confidence regions of a two-dimensional Gaussian distribution. The star indicates the values of $S_8$ and $\Omega_\mathrm{m}$ in the UNIT simulations. Finally, small grey dots indicate the cosmological parameters of AbacusSummit simulations used in this analysis.}
    \label{fig:mock_cosmo}
\end{figure}

In Fig.~\ref{fig:mock_best_fit}, we show the best-fit galaxy model on the best-fit AbacusSummit simulation. We see that our galaxy model can fit the mock data vector well with $\chi^2 = 10.6$ ($\chi_\nu^2 = 10.6 / (18 - 6.8 - 2)$, $p=0.32$). In Fig.~\ref{fig:mock_cosmo}, we show the cosmological constraints in addition to the input cosmology of the UNIT simulations. We find that the UNIT cosmology is at $\Delta \chi^2 = 3.4$ from the best-fit cosmology. For reference, for a two-dimensional Gaussian, $19\%$ of random draws will result in a higher $\Delta \chi^2$, indicating that our cosmological constraints are consistent with being unbiased. Nonetheless, we stress the need for additional mock tests in the future, especially for the BGS sample which may additionally be affected by the limited resolution of the AbacusSummit base simulations.

\subsection{Masking}

We aim to protect analysis choices in this work from confirmation bias and other subconscious effects that could bias the outcome of our analysis. To this end, we first perform ``masked'' analysis on the DESI data that allows us to implement quality controls, such as checking the goodness of fit, while being insensitive to the constraints on $\Omega_\mathrm{m}$ and $S_8$. We implement a data vector masking, similar to \cite{Muir2020_MNRAS_494_4454} and \cite{Lange2023_MNRAS_520_5373}. We find that simply multiplying the lensing amplitude $\Delta\Sigma$ by an unknown factor $A_\mathrm{lens}$, while keeping $w_\mathrm{p}$ unchanged, moves constraints mainly along the $S_8$ line while having only a small impact on the goodness of fit. Thus, to create the masked observational data, we draw $A_\mathrm{lens}$ from a uniform distribution in the range $[0.85, 1.15]$ and multiply the lensing amplitudes for all DESI samples and lensing measurements by the same $A_\mathrm{lens}$.

When working with the masked data, based on the goodness-of-fit, we determined that the mass resolution of the AbacusSummit simulations was insufficient for BGS1. No other significant changes were made in response to analyzing the masked data.

\section{Results}
\label{sec:results}

Here, we present the unmasked cosmology results. The results of this analysis were only unmasked after it was determined that the galaxy model can obtain an acceptable fit to the data and that the different DESI samples, BGS2, BGS3, LRG1, and LRG2, give consistent cosmological constraints. The only significant post-unmasking change was fixing an error where correlations between lensing measurements of the same lenses but different source tomographic bins were neglected. That caused uncertainties for the lensing measurements and thereby $S_8$ uncertainties to be underestimated by $\sim 10\%$ to $20 \%$.

\subsection{Goodness of Fit}

\begin{figure*}
    \centering
    \includegraphics{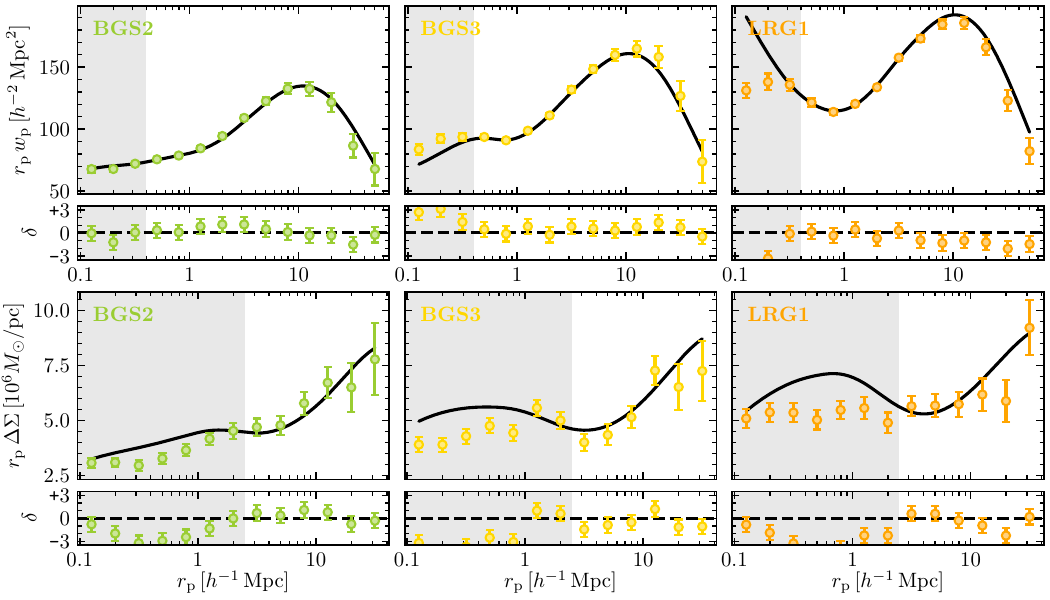}
    \caption{Similar to Fig~\ref{fig:mock_best_fit} but for the DESI measurements with DES and KiDS lensing. The best-fit simulation, \texttt{AbacusSummit\_base\_c103\_ph000}, is the one that minimizes the total $\chi^2$ of all three samples whereas the galaxy model is optimized individually for each sample.}
    \label{fig:data_best_fit}
\end{figure*}

\begin{figure*}
    \centering
    \includegraphics{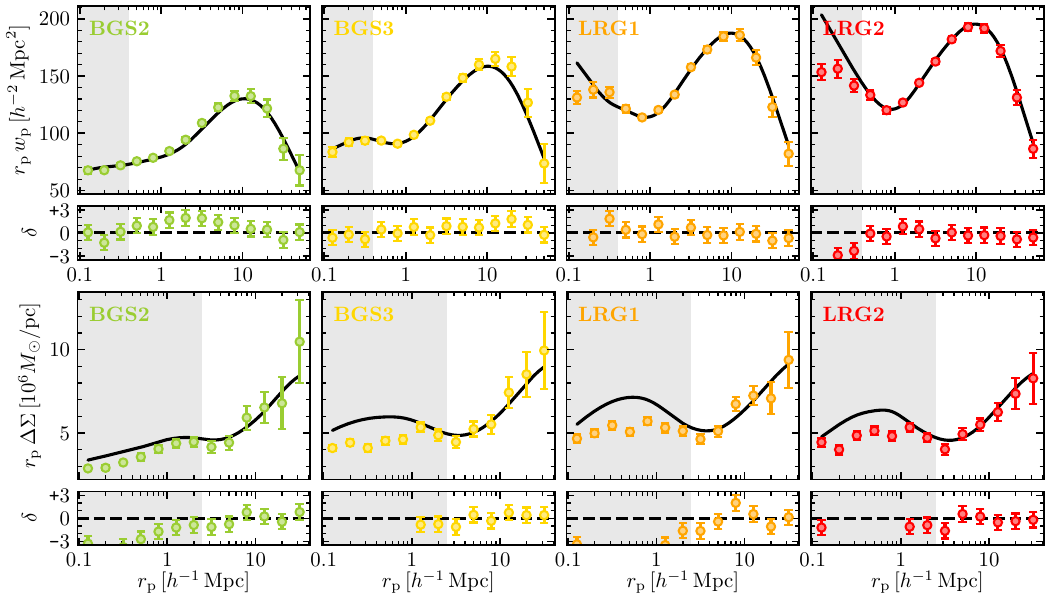}
    \caption{Similar to Fig~\ref{fig:data_best_fit} but for the DESI measurements with HSC lensing amplitudes. The best-fitting simulation is \texttt{AbacusSummit\_base\_c118\_ph000}.}
    \label{fig:data_best_fit_hsc}
\end{figure*}

\begin{figure*}
    \centering
    \includegraphics{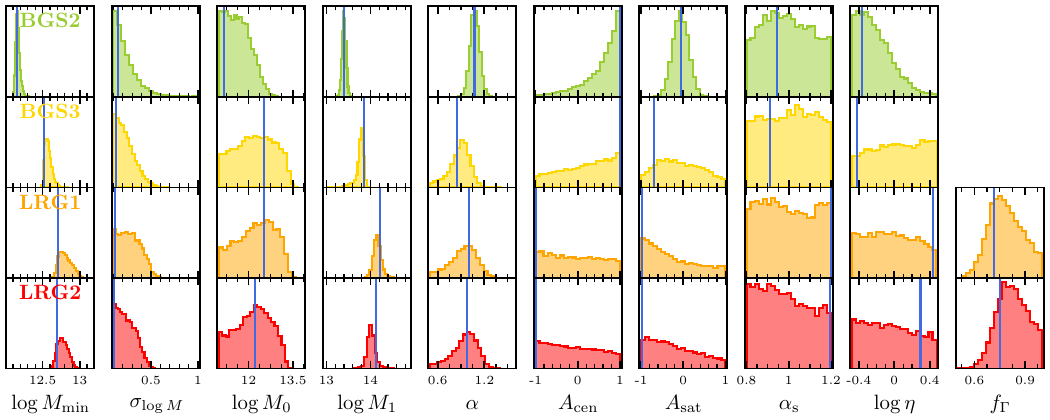}
    \caption{One-dimensional Bayesian posteriors on the galaxy model parameters for the four DESI samples with HSC lensing amplitudes under the best-fit cosmology. We also show the best-fit parameters as blue vertical lines. The different parameters are described in \protect\ref{tab:galaxy_parameters}.}
    \label{fig:data_hod_posterior}
\end{figure*}

In Fig.~\ref{fig:data_best_fit}, we show the best-fit model predictions for the different DESI samples studied in this work with lensing amplitudes from DES/KiDS. Fig.~\ref{fig:data_best_fit_hsc} shows a similar result but for lensing amplitudes from HSC. We find that we can obtain good fits to the data of all DESI lens samples with a single simulation, i.e., cosmology. The total goodness of fit is $\chi^2 = 48.5$ with $54$ data points for DES/KiDS and $\chi^2 = 55.1$ with $72$ data points for HSC. We have two degrees of freedom for the cosmology fit but the number of effective degrees of freedom of the galaxy fit is significantly less than the number of galaxy model parameters since many are effectively unconstrained, as shown in Fig.~\ref{fig:data_hod_posterior}. We estimate the effective degrees of freedom numerically. We start with the best-fit galaxy model $\theta_0$ for each sample at AbacusSummit cosmology $0$. We then repeatedly draw mock observations from a multivariate normal distribution with the predictions from $\theta_0$ as the mean and the observational uncertainties as the covariance. Finally, we minimize the $\chi^2$ over the galaxy parameters and calculate the difference between the minimum $\chi^2$ and $\chi^2(\theta_0)$. Based on the reduction in the $\chi^2$, we estimate the effective degrees of freedom in the galaxy parameters over all DESI samples to be around $14.0$ for the DES/KiDS fits and $20.3$ for the HSC fits. This gives $\chi_\nu^2 = 48.5 / (54 - 2 - 14.0) $ ($p=0.12$) for DES/KiDS and $\chi_\nu^2 = 55.1 / (72 - 2 - 20.3)$ ($p=0.28$) for HSC. Overall, this indicates an acceptable fit of our $\Lambda$CDM and HOD-based model to the observational data.

\subsection{Cosmology}

\begin{figure*}
    \centering
    \includegraphics{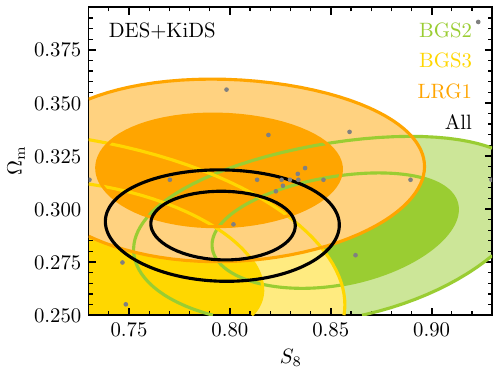}
    \includegraphics{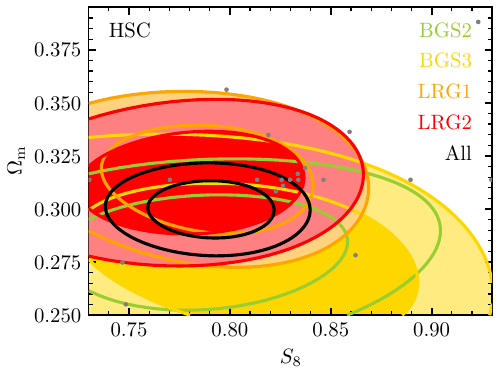}
    \caption{Observational constraints on $S_8$ and $\Omega_\mathrm{m}$ for DESI clustering measurements combined with DES Y3 and KiDS-1000 lensing measurements (left) and HSC Y3 lensing amplitudes (right). The different shaded regions demarcate the $\Delta\chi^2 = 2.28$ and $\Delta\chi^2 = 5.99$ boundaries. The black contour indicates the combined results by adding the $\chi^2$ contributions from the individual samples. As in Fig.~\protect\ref{fig:mock_cosmo}, small grey dots indicate the AbacusSummit simulations.}
    \label{fig:data_cosmo}
\end{figure*}

The main result of our analysis, observational constraints on $S_8$ and $\Omega_\mathrm{m}$, are shown in Fig.~\ref{fig:data_cosmo}. We show the individual constraints from the different samples as well as their combination. The best-fit $\chi^2$ when fitting all samples with a single combination of $S_8$ and $\Omega_\mathrm{m}$ increases by $\Delta\chi^2 = 7.0$ for DES/KiDS and $3.7$ for HSC compared to optimizing those two parameters for each individual sample. This increase in $\chi^2$ is consistent with reducing the degrees of freedom by $4$ for DES/KiDS ($\Delta \chi_\nu^2 = 7.0 / 4$, $p = 0.14$) and $6$ for HSC ($\Delta \chi_\nu^2 = 3.7 / 6$, $p = 0.72$), indicating that the different DESI samples produce consistent results on $S_8$ and $\Omega_\mathrm{m}$. Our final marginalized (Frequentist) constraints are $S_8 = 0.797_{-0.024}^{+0.023}$ and $\Omega_\mathrm{m} = 0.292 \pm 0.011$ for DES/KiDS and $S_8 = 0.791_{-0.021}^{+0.020}$ and $\Omega_\mathrm{m} = 0.300 \pm 0.009$ for HSC. Estimates correspond to the maximum-likelihood estimate (MLE) whereas uncertainties correspond to $\Delta \chi^2 = 1$ from the MLE, i.e., $68\%$ confidence intervals assuming a one-dimensional normal distribution. As expected given the consistency in the measured $\Delta\Sigma$ amplitudes \citep{Heydenreich2025_arXiv_2506_1677}, we see consistency between the DES+KiDS and HSC constraints on $S_8$ and $\Omega_\mathrm{m}$.

For HSC, the constraints listed above assume the best-fit photometric redshift offsets found by \cite{Li2023_PhRvD_108_3518}. If we assume no offsets, we find $S_8 = 0.841 \pm 0.020$, a $\sim 2.5 \sigma$ shift, and $\Omega_\mathrm{m} = 0.305 \pm 0.009$, instead. However, the total $\chi^2$ increases by $7.1$, indicating that this scenario is disfavored by the data analyzed in this work. Given the shift of $\Delta S_8 = 0.050$ with $\Delta \chi^2 = 7.1$, we estimate that a full marginalization over photometric redshift calibration uncertainties would roughly add $\sim \Delta S_8 / \sqrt{\Delta \chi^2} = 0.019$ to the overall uncertainty in $S_8$. In other words, the results for HSC are likely limited by systematic uncertainties in the photometric redshift calibration. Since unmasking of the cosmology constraints of this work, more recent studies \citep[][Fisher et al. in preparation]{ChoppindeJanvry2026_JCAP_05_073} suggest that photometric redshift offsets are likely significantly smaller than the best-fit values by \cite{Li2023_PhRvD_108_3518}, which would lead to higher inferred $S_8$ values. Future studies analyzing DESI and HSC data jointly will employ these updated calibrations and significantly reduce systematic uncertainties related to photometric redshifts.

\begin{figure}
    \centering
    \includegraphics{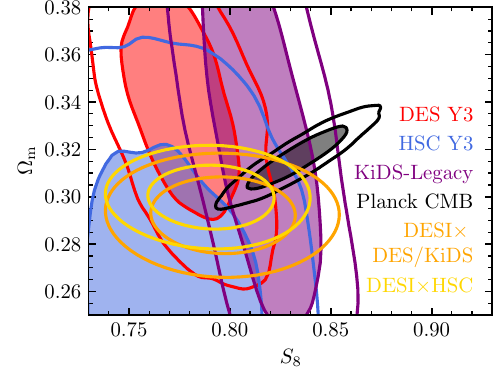}
    \caption{Comparison of the constraints derived in this work against the \cite{PlanckCollaboration2020_AA_641_6} TT,TE,EE+lowE CMB results and the latest publicly available cosmology results from DES Y3 \citep{Abbott2022_PhRvD_105_3520}, HSC Y3 \citep{Li2023_PhRvD_108_3518}, and KiDS-Legacy \citep{Wright2025_AA_703_158}. Filled contours and non-filled contours show the $68\%$ and $95\%$ containment ranges, respectively. For visual clarity, each of the publicly available results have been smoothed with a Gaussian kernel that has a covariance that is $4\%$ that of the posterior.}
    \label{fig:data_cosmo_vs_cmb}
\end{figure}

We compare our results derived here against the \cite{PlanckCollaboration2020_AA_641_6} CMB results as well as recent weak lensing results from DES, HSC, and KiDS in Fig.~\ref{fig:data_cosmo_vs_cmb}. Overall, our results are consistent with the \cite{PlanckCollaboration2020_AA_641_6} constraints albeit preferring a lower value for $S_8$ at $1.5$ (DES+KiDS) to $1.8 \sigma$ (HSC). Consequently, our results are also consistent with recent weak lensing results which tend to have a preference for lower $S_8$ than CMB constraints.

Finally, it is worth comparing our results from the combination of clustering and galaxy--galaxy lensing to those derived in \citep{Porredon2026_OJAp_9_1342}. The main difference compared to our work is a different choice of scale cuts: $8 \, h^{-1} \, \mathrm{Mpc}$ for clustering and $6 \, h^{-1} \, \mathrm{Mpc}$. Additionally, \cite{Porredon2026_OJAp_9_1342} use point-mass marginalization which further reduces the sensitivity to the lensing measurements near the scale cut \citep{Prat2022_PhRvD_105_3528}. Ultimately, \cite{Porredon2026_OJAp_9_1342} find $S_8 = 0.707_{-0.086}^{+0.090}$, $0.707_{-0.098}^{+0.094}$, and $0.649_{-0.067}^{+0.091}$ for DES, KiDS, and HSC, respectively. These constraints are substantially weaker than those derived in this work, highlighting the constraining power of non-linear scales.

\section{Conclusion}
\label{sec:conclusion}

In this work, we present new constraints of cosmic structure growth from the analysis of non-linear clustering and galaxy-galaxy lensing with DESI DR1. We implement a complex, simulation-based modeling framework that accounts for important details in the galaxy-halo connection such as galaxy assembly bias. Overall, our results align well with previous literature results. In particular, comparing with \cite{Porredon2026_OJAp_9_1342}, we find that the analysis of non-linear scales substantially improves the cosmological constraining power of the data \citep[see, e.g.,][]{Reid2014_MNRAS_444_476, Wibking2019_MNRAS_484_989, Lange2022_MNRAS_509_1779}.

Our analysis places stringent constraints on cosmic structure growth, particularly the $S_8$ parameter. While our maximum-likelihood value for $S_8$ falls below that of the \cite{PlanckCollaboration2020_AA_641_6} CMB analysis, similar to many other low-redshift studies \citep[see, e.g.,][]{Abdalla2022_JHEAp_34_49, Li2023_PhRvD_108_3518, Wright2025_AA_703_158, DESCollaboration2026_arXiv_2601_4559}, our results are also not in significant tension with CMB results. This finding is similar to other recent studies of non-linear clustering and lensing \citep{Wibking2020_MNRAS_492_2872, Dvornik2023_AA_675_189, Amon2023_MNRAS_518_477, Lange2023_MNRAS_520_5373}. Our results also align with earlier studies reporting the so-called lensing-is-low problem \citep{Leauthaud2017_MNRAS_467_3024, Lange2019_MNRAS_488_5771}. This problem arises when comparing the lensing amplitude on very small scales, down to $0.1 \, h^{-1} \, \mathrm{Mpc}$, to predictions of simple mass-only HOD models and without the expected impact of baryons. In this work, we take into account assembly bias and do not model the scales expected to be most strongly affected by baryonic physics. Indeed, looking at Figs.~\ref{fig:data_best_fit} and \ref{fig:data_best_fit_hsc}, we see that our best-fit models consistently overpredict lensing amplitude below $1 \, h^{-1} \, \mathrm{Mpc}$.

The present work serves as a precursor for future studies with DESI data that promise further increases in statistical constraining power. To fully realize this potential, further steps need to be taken to reduce systematic and modeling uncertainties. In the current analysis, we exclude the lensing signal on scales below $2.5 \, h^{-1} \, \mathrm{Mpc}$ from the analysis due to the potential impact of baryonic physics. If we were to constrain baryonic feedback, we could potentially make use of the high-precision lensing measurements on these very small scales. Similarly, despite advances in recent years, further mock tests should be conducted to investigate the robustness of cosmological constraints from non-linear scales \citep{Doytcheva2024_ApJ_977_184, ThBeyond2ptCollaboration2025_ApJ_990_99, Zhai2025_MNRAS_544_932}. Finally, specifically for gravitational lensing with HSC, reducing uncertainties in the photometric redshift calibration is critical.

In the future, joint studies of clustering, lensing, and the Sunyaev-Zel'dovich effect could significantly reduce the uncertainty on baryonic physics and allow us to model smaller scales \citep{Amodeo2021_PhRvD_103_3514, Bigwood2024_MNRAS_534_655, Sunseri2026_MNRAS_546_g031}. In particular, the overlap of DESI with the footprint of the Atacama Cosmology Telescope (ACT) enables high-precision measurements of the kinetic Sunyaev-Zel'dovich effect \citep{Hadzhiyska2025_PhRvD_112_3509}.
Similarly, DESI is an excellent source for high-quality spectroscopic redshifts that will allow us to reduce uncertainties in the HSC photometric redshift calibration. This can be done both via clustering redshifts using DESI large-scale structure catalogs \citep{ChoppindeJanvry2026_JCAP_05_073, Ruggeri2025_arXiv_2512_5963} and direct calibration using DESI redshifts in deep drilling fields \citep{Ratajczak2026_AJ_171_71, Lange2026_ApJ_998_263}.

\section*{Data Availability}

All data points shown in the published graphs will be available at \url{https://zenodo.org/records/17831718} upon publication. We have released all {\sc TabCorr} data products used in this work to the community at \url{https://zenodo.org/records/15588541} as part of the {\sc TabCorr} library.

\section*{Acknowledgments}

We acknowledge use of the lux supercomputer at UC Santa Cruz, funded by NSF MRI grant AST 1828315. During the revision of this manuscript, \textsc{Claude Sonnet 5} was used to check parts of the code used in this analysis for errors and inconsistencies with the text. No changes to the code or text were made as a result. All text and code for this work was produced by the authors.

\textbf{Dark Energy Spectroscopic Instrument}:  This material is based upon work supported by the U.S. Department of Energy (DOE), Office of Science, Office of High-Energy Physics, under Contract No. DE–AC02–05CH11231, and by the National Energy Research Scientific Computing Center, a DOE Office of Science User Facility under the same contract. Additional support for DESI was provided by the U.S. National Science Foundation (NSF), Division of Astronomical Sciences under Contract No. AST-0950945 to the NSF’s National Optical-Infrared Astronomy Research Laboratory; the Science and Technology Facilities Council of the United Kingdom; the Gordon and Betty Moore Foundation; the Heising-Simons Foundation; the French Alternative Energies and Atomic Energy Commission (CEA); the National Council of Humanities, Science and Technology of Mexico (CONAHCYT); the Ministry of Science, Innovation and Universities of Spain (MICIU/AEI/10.13039/501100011033), and by the DESI Member Institutions: \url{https://www.desi.lbl.gov/collaborating-institutions}. Any opinions, findings, and conclusions or recommendations expressed in this material are those of the author(s) and do not necessarily reflect the views of the U. S. National Science Foundation, the U. S. Department of Energy, or any of the listed funding agencies.

The authors are honored to be permitted to conduct scientific research on I'oligam Du'ag (Kitt Peak), a mountain with particular significance to the Tohono O’odham Nation.

\textbf{Dark Energy Survey}: This project used public archival data from the Dark Energy Survey (DES). Funding for the DES Projects has been provided by the U.S. Department of Energy, the U.S. National Science Foundation, the Ministry of Science and Education of Spain, the Science and Technology FacilitiesCouncil of the United Kingdom, the Higher Education Funding Council for England, the National Center for Supercomputing Applications at the University of Illinois at Urbana-Champaign, the Kavli Institute of Cosmological Physics at the University of Chicago, the Center for Cosmology and Astro-Particle Physics at the Ohio State University, the Mitchell Institute for Fundamental Physics and Astronomy at Texas A\&M University, Financiadora de Estudos e Projetos, Funda{\c c}{\~a}o Carlos Chagas Filho de Amparo {\`a} Pesquisa do Estado do Rio de Janeiro, Conselho Nacional de Desenvolvimento Cient{\'i}fico e Tecnol{\'o}gico and the Minist{\'e}rio da Ci{\^e}ncia, Tecnologia e Inova{\c c}{\~a}o, the Deutsche Forschungsgemeinschaft, and the Collaborating Institutions in the Dark Energy Survey.

The Collaborating Institutions are Argonne National Laboratory, the University of California at Santa Cruz, the University of Cambridge, Centro de Investigaciones Energ{\'e}ticas, Medioambientales y Tecnol{\'o}gicas-Madrid, the University of Chicago, University College London, the DES-Brazil Consortium, the University of Edinburgh, the Eidgen{\"o}ssische Technische Hochschule (ETH) Z{\"u}rich,  Fermi National Accelerator Laboratory, the University of Illinois at Urbana-Champaign, the Institut de Ci{\`e}ncies de l'Espai (IEEC/CSIC), the Institut de F{\'i}sica d'Altes Energies, Lawrence Berkeley National Laboratory, the Ludwig-Maximilians Universit{\"a}t M{\"u}nchen and the associated Excellence Cluster Universe, the University of Michigan, the National Optical Astronomy Observatory, the University of Nottingham, The Ohio State University, the OzDES Membership Consortium, the University of Pennsylvania, the University of Portsmouth, SLAC National Accelerator Laboratory, Stanford University, the University of Sussex, and Texas A\&M University.

Based in part on observations at Cerro Tololo Inter-American Observatory, National Optical Astronomy Observatory, which is operated by the Association of Universities for Research in Astronomy (AURA) under a cooperative agreement with the National Science Foundation.

\textbf{Hyper Suprime-Cam}: The Hyper Suprime-Cam (HSC) collaboration includes the astronomical communities of Japan and Taiwan, and Princeton University. The HSC instrumentation and software were developed by the National Astronomical Observatory of Japan (NAOJ), the Kavli Institute for the Physics and Mathematics of the Universe (Kavli IPMU), the University of Tokyo, the High Energy Accelerator Research Organization (KEK), the Academia Sinica Institute for Astronomy and Astrophysics in Taiwan (ASIAA), and Princeton University. Funding was contributed by the FIRST program from the Japanese Cabinet Office, the Ministry of Education, Culture, Sports, Science and Technology (MEXT), the Japan Society for the Promotion of Science (JSPS), Japan Science and Technology Agency (JST), the Toray Science Foundation, NAOJ, Kavli IPMU, KEK, ASIAA, and Princeton University. 

This paper makes use of software developed for Vera C. Rubin Observatory. We thank the Rubin Observatory for making their code available as free software at \url{http://pipelines.lsst.io/}.

This paper is based on data collected at the Subaru Telescope and retrieved from the HSC data archive system, which is operated by the Subaru Telescope and Astronomy Data Center (ADC) at NAOJ. Data analysis was in part carried out with the cooperation of Center for Computational Astrophysics (CfCA), NAOJ. We are honored and grateful for the opportunity of observing the Universe from Maunakea, which has the cultural, historical and natural significance in Hawaii. 

\textbf{Kilo Degree-Survey}: Based on observations made with ESO Telescopes at the La Silla Paranal Observatory under programme IDs 177.A-3016, 177.A-3017, 177.A-3018 and 179.A-2004, and on data products produced by the KiDS consortium. The KiDS production team acknowledges support from: Deutsche Forschungsgemeinschaft, ERC, NOVA and NWO-M grants; Target; the University of Padova, and the University Federico II (Naples).

We use the gold sample of weak lensing and photometric redshift measurements from the fourth data release of the Kilo-Degree Survey \citep{Kuijken2019_AA_625_2, Wright2020_AA_637_100, Hildebrandt2021_AA_647_124, Giblin2021_AA_645_105}, hereafter referred to as KiDS-1000. Cosmological parameter constraints from KiDS-1000 have been presented in \cite[][cosmic shear]{Asgari2021_AA_645_104}, \cite[][$3\times2$pt]{Heymans2021_AA_646_140} and \cite[][beyond $\Lambda$CDM]{Troster2021_AA_649_88}, with the methodology presented in \cite{Joachimi2021_AA_646_129}.

\facilities{Mayall, Blanco, Subaru, ESO:VISTA, VLT}

\software{{\sc Astropy} \citep{AstropyCollaboration2013_AA_558_33, AstropyCollaboration2018_AJ_156_123, AstropyCollaboration2022_ApJ_935_167}, {\sc Corrfunc} \citep{Sinha2020_MNRAS_491_3022}, {\sc dsigma} \citep{Lange2022_ascl_soft_4006} {\sc halotools} \citep{Hearin2017_AJ_154_190}, {\sc matplotlib} \citep{Hunter2007_CSE_9_90}, {\sc NumPy} \citep{VanDerWalt2011_CSE_13_22}, {\sc SciPy} \citep{Virtanen2020_NaMet_17_261}, {\sc TabCorr} \citep{Lange2019_MNRAS_490_1870}, {\sc Spyder}, \textsc{Claude Sonnet 5}}

\bibliography{bibliography}
\bibliographystyle{aasjournalv7}

\end{document}

%% file: authors.tex
\author[0000-0002-2450-1366]{J.~U.~Lange}
\email{jlange@american.edu}
\affiliation{Department of Physics, American University, 4400 Massachusetts Avenue NW, Washington, DC 20016, USA}

\author[0000-0001-5080-4582]{A.~Wells}
\email{acwells@umich.edu}
\affiliation{Department of Physics, University of Michigan, 450 Church Street, Ann Arbor, MI 48109, USA}

\author{A.~Hearin}
\email{ahearin@anl.gov}
\affiliation{Argonne National Laboratory, High-Energy Physics Division, 9700 S. Cass Avenue, Argonne, IL 60439, USA}

\author[0000-0002-4392-8920]{G.~Beltz-Mohrmann}
\email{gbeltzmohrmann@smith.edu}
\affiliation{Department of Physics, Smith College, 1 Chapin Way, Northampton, MA 01063, USA}

\author[0000-0002-3677-3617]{A.~Leauthaud}
\email{alexie@ucsc.edu}
\affiliation{Department of Astronomy and Astrophysics, UCO/Lick Observatory, University of California, 1156 High Street, Santa Cruz, CA 95064, USA}
\affiliation{Department of Astronomy and Astrophysics, University of California, Santa Cruz, 1156 High Street, Santa Cruz, CA 95065, USA}

\author[0000-0002-7273-4076]{S.~Heydenreich}
\email{sheydenr@ucsc.edu}
\affiliation{Department of Astronomy and Astrophysics, UCO/Lick Observatory, University of California, 1156 High Street, Santa Cruz, CA 95064, USA}

\author[0000-0002-5423-5919]{C.~Blake}
\email{cblake@astro.swin.edu.au}
\affiliation{Centre for Astrophysics \& Supercomputing, Swinburne University of Technology, P.O. Box 218, Hawthorn, VIC 3122, Australia}

\author{J.~Aguilar}
\email{jaguilar@lbl.gov}
\affiliation{Lawrence Berkeley National Laboratory, 1 Cyclotron Road, Berkeley, CA 94720, USA}

\author[0000-0001-6098-7247]{S.~Ahlen}
\email{ahlen@bu.edu}
\affiliation{Department of Physics, Boston University, 590 Commonwealth Avenue, Boston, MA 02215 USA}

\author[0000-0003-2923-1585]{A.~Anand}
\email{abhijeetanand2011@gmail.com}
\affiliation{Lawrence Berkeley National Laboratory, 1 Cyclotron Road, Berkeley, CA 94720, USA}

\author[0000-0001-9712-0006]{D.~Bianchi}
\email{davide.bianchi1@unimi.it}
\affiliation{Dipartimento di Fisica ``Aldo Pontremoli'', Universit\`a degli Studi di Milano, Via Celoria 16, I-20133 Milano, Italy}
\affiliation{INAF-Osservatorio Astronomico di Brera, Via Brera 28, 20122 Milano, Italy}

\author{D.~Brooks}
\email{david.brooks@ucl.ac.uk}
\affiliation{Department of Physics \& Astronomy, University College London, Gower Street, London, WC1E 6BT, UK}

\author[0000-0001-7316-4573]{F.~J.~Castander}
\email{fjc@ice.csic.es}
\affiliation{Institut d'Estudis Espacials de Catalunya (IEEC), c/ Esteve Terradas 1, Edifici RDIT, Campus PMT-UPC, 08860 Castelldefels, Spain}
\affiliation{Institute of Space Sciences, ICE-CSIC, Campus UAB, Carrer de Can Magrans s/n, 08913 Bellaterra, Barcelona, Spain}

\author{T.~Claybaugh}
\email{tmclaybaugh@lbl.gov}
\affiliation{Lawrence Berkeley National Laboratory, 1 Cyclotron Road, Berkeley, CA 94720, USA}

\author[0000-0002-5954-7903]{S.~Cole}
\email{shaun.cole@durham.ac.uk}
\affiliation{Institute for Computational Cosmology, Department of Physics, Durham University, South Road, Durham DH1 3LE, UK}

\author[0000-0002-2169-0595]{A.~Cuceu}
\email{acuceu@lbl.gov}
\affiliation{Lawrence Berkeley National Laboratory, 1 Cyclotron Road, Berkeley, CA 94720, USA}

\author[0000-0002-0553-3805]{K.~S.~Dawson}
\email{kdawson@astro.utah.edu}
\affiliation{Department of Physics and Astronomy, The University of Utah, 115 South 1400 East, Salt Lake City, UT 84112, USA}

\author[0000-0002-1769-1640]{A.~de la Macorra}
\email{macorra@fisica.unam.mx}
\affiliation{Instituto de F\'{\i}sica, Universidad Nacional Aut\'{o}noma de M\'{e}xico,  Circuito de la Investigaci\'{o}n Cient\'{\i}fica, Ciudad Universitaria, Cd. de M\'{e}xico  C.~P.~04510,  M\'{e}xico}

\author[0000-0002-5665-7912]{Biprateep~Dey}
\email{b.dey@utoronto.ca}
\affiliation{Department of Astronomy \& Astrophysics, University of Toronto, Toronto, ON M5S 3H4, Canada}
\affiliation{Department of Physics \& Astronomy and Pittsburgh Particle Physics, Astrophysics, and Cosmology Center (PITT PACC), University of Pittsburgh, 3941 O'Hara Street, Pittsburgh, PA 15260, USA}

\author{P.~Doel}
\email{apd@star.ucl.ac.uk}
\affiliation{Department of Physics \& Astronomy, University College London, Gower Street, London, WC1E 6BT, UK}

\author[0000-0001-6537-6453]{A.~Elliott}
\email{elliott.222@osu.edu}
\affiliation{Department of Physics, The Ohio State University, 191 West Woodruff Avenue, Columbus, OH 43210, USA}
\affiliation{The Ohio State University, Columbus, 43210 OH, USA}

\author[0000-0001-8097-6019]{N.~Emas}
\email{nemas@swin.edu.au}
\affiliation{Centre for Astrophysics \& Supercomputing, Swinburne University of Technology, P.O. Box 218, Hawthorn, VIC 3122, Australia}

\author[0000-0003-4992-7854]{S.~Ferraro}
\email{sferraro@lbl.gov}
\affiliation{Lawrence Berkeley National Laboratory, 1 Cyclotron Road, Berkeley, CA 94720, USA}
\affiliation{University of California, Berkeley, 110 Sproul Hall \#5800 Berkeley, CA 94720, USA}

\author[0000-0002-3033-7312]{A.~Font-Ribera}
\email{afont@ifae.es}
\affiliation{Instituci\'{o} Catalana de Recerca i Estudis Avan\c{c}ats, Passeig de Llu\'{\i}s Companys, 23, 08010 Barcelona, Spain}
\affiliation{Institut de F\'{i}sica d’Altes Energies (IFAE), The Barcelona Institute of Science and Technology, Edifici Cn, Campus UAB, 08193, Bellaterra (Barcelona), Spain}

\author[0000-0002-2890-3725]{J.~E.~Forero-Romero}
\email{je.forero@uniandes.edu.co}
\affiliation{Departamento de F\'isica, Universidad de los Andes, Cra. 1 No. 18A-10, Edificio Ip, CP 111711, Bogot\'a, Colombia}
\affiliation{Observatorio Astron\'omico, Universidad de los Andes, Cra. 1 No. 18A-10, Edificio H, CP 111711 Bogot\'a, Colombia}

\author[0000-0003-1481-4294]{C.~Garcia-Quintero}
\email{cristhian.garcia_quintero@cfa.harvard.edu}
\affiliation{Center for Astrophysics $|$ Harvard \& Smithsonian, 60 Garden Street, Cambridge, MA 02138, USA}
\affiliation{NASA Einstein Fellow}

\author[0000-0001-9632-0815]{E.~Gaztañaga}
\email{gaztanaga@gmail.com}
\affiliation{Institut d'Estudis Espacials de Catalunya (IEEC), c/ Esteve Terradas 1, Edifici RDIT, Campus PMT-UPC, 08860 Castelldefels, Spain}
\affiliation{Institute of Cosmology and Gravitation, University of Portsmouth, Dennis Sciama Building, Portsmouth, PO1 3FX, UK}
\affiliation{Institute of Space Sciences, ICE-CSIC, Campus UAB, Carrer de Can Magrans s/n, 08913 Bellaterra, Barcelona, Spain}

\author[0000-0003-3142-233X]{Satya~{Gontcho A Gontcho}}
\email{satya@virginia.edu}
\affiliation{University of Virginia, Department of Astronomy, Charlottesville, VA 22904, USA}

\author{G.~Gutierrez}
\email{gaston@fnal.gov}
\affiliation{Fermi National Accelerator Laboratory, PO Box 500, Batavia, IL 60510, USA}

\author[0000-0001-9822-6793]{J.~Guy}
\email{jguy@lbl.gov}
\affiliation{Lawrence Berkeley National Laboratory, 1 Cyclotron Road, Berkeley, CA 94720, USA}

\author[0000-0002-6550-2023]{K.~Honscheid}
\email{kh@physics.osu.edu}
\affiliation{Center for Cosmology and AstroParticle Physics, The Ohio State University, 191 West Woodruff Avenue, Columbus, OH 43210, USA}
\affiliation{Department of Physics, The Ohio State University, 191 West Woodruff Avenue, Columbus, OH 43210, USA}
\affiliation{The Ohio State University, Columbus, 43210 OH, USA}

\author[0000-0001-6558-0112]{D.~Huterer}
\email{huterer@umich.edu}
\affiliation{Department of Physics, University of Michigan, 450 Church Street, Ann Arbor, MI 48109, USA}
\affiliation{University of Michigan, 500 S. State Street, Ann Arbor, MI 48109, USA}

\author[0000-0002-6024-466X]{M.~Ishak}
\email{mishak@utdallas.edu}
\affiliation{Department of Physics, The University of Texas at Dallas, 800 W. Campbell Rd., Richardson, TX 75080, USA}

\author[0000-0001-8820-673X]{S.~Joudaki}
\email{shahab.joudaki@ciemat.es}
\affiliation{CIEMAT, Avenida Complutense 40, E-28040 Madrid, Spain}

\author[0000-0003-0201-5241]{R.~Joyce}
\email{richard.joyce@noirlab.edu}
\affiliation{NSF NOIRLab, 950 N. Cherry Ave., Tucson, AZ 85719, USA}

\author{R.~Kehoe}
\email{kehoe@physics.smu.edu}
\affiliation{Department of Physics, Southern Methodist University, 3215 Daniel Avenue, Dallas, TX 75275, USA}

\author[0000-0002-8828-5463]{D.~Kirkby}
\email{dkirkby@uci.edu}
\affiliation{Department of Physics and Astronomy, University of California, Irvine, 92697, USA}

\author[0000-0003-3510-7134]{T.~Kisner}
\email{tskisner@lbl.gov}
\affiliation{Lawrence Berkeley National Laboratory, 1 Cyclotron Road, Berkeley, CA 94720, USA}

\author[0000-0001-6356-7424]{A.~Kremin}
\email{akremin@lbl.gov}
\affiliation{Lawrence Berkeley National Laboratory, 1 Cyclotron Road, Berkeley, CA 94720, USA}

\author{A.~Krolewski}
\email{akrolews@uwaterloo.ca}
\affiliation{Department of Physics and Astronomy, University of Waterloo, 200 University Ave W, Waterloo, ON N2L 3G1, Canada}
\affiliation{Perimeter Institute for Theoretical Physics, 31 Caroline St. North, Waterloo, ON N2L 2Y5, Canada}
\affiliation{Waterloo Centre for Astrophysics, University of Waterloo, 200 University Ave W, Waterloo, ON N2L 3G1, Canada}

\author[0000-0002-1134-9035]{O.~Lahav}
\email{o.lahav@ucl.ac.uk}
\affiliation{Department of Physics \& Astronomy, University College London, Gower Street, London, WC1E 6BT, UK}

\author[0000-0002-6731-9329]{C.~Lamman}
\email{lamman.1@osu.edu}
\affiliation{The Ohio State University, Columbus, 43210 OH, USA}

\author[0000-0003-1838-8528]{M.~Landriau}
\email{mlandriau@lbl.gov}
\affiliation{Lawrence Berkeley National Laboratory, 1 Cyclotron Road, Berkeley, CA 94720, USA}

\author[0000-0001-7178-8868]{L.~Le~Guillou}
\email{llg@lpnhe.in2p3.fr}
\affiliation{Sorbonne Universit\'{e}, CNRS/IN2P3, Laboratoire de Physique Nucl\'{e}aire et de Hautes Energies (LPNHE), FR-75005 Paris, France}

\author[0000-0003-1887-1018]{M.~E.~Levi}
\email{melevi@lbl.gov}
\affiliation{Lawrence Berkeley National Laboratory, 1 Cyclotron Road, Berkeley, CA 94720, USA}

\author[0000-0003-4962-8934]{M.~Manera}
\email{mmanera@ifae.es}
\affiliation{Departament de F\'{i}sica, Serra H\'{u}nter, Universitat Aut\`{o}noma de Barcelona, 08193 Bellaterra (Barcelona), Spain}
\affiliation{Institut de F\'{i}sica d’Altes Energies (IFAE), The Barcelona Institute of Science and Technology, Edifici Cn, Campus UAB, 08193, Bellaterra (Barcelona), Spain}

\author[0000-0002-4279-4182]{P.~Martini}
\email{martini.10@osu.edu}
\affiliation{Center for Cosmology and AstroParticle Physics, The Ohio State University, 191 West Woodruff Avenue, Columbus, OH 43210, USA}
\affiliation{Department of Astronomy, The Ohio State University, 4055 McPherson Laboratory, 140 W 18th Avenue, Columbus, OH 43210, USA}
\affiliation{The Ohio State University, Columbus, 43210 OH, USA}

\author[0000-0002-1125-7384]{A.~Meisner}
\email{aaron.meisner@noirlab.edu}
\affiliation{NSF NOIRLab, 950 N. Cherry Ave., Tucson, AZ 85719, USA}

\author{R.~Miquel}
\email{rmiquel@ifae.es}
\affiliation{Instituci\'{o} Catalana de Recerca i Estudis Avan\c{c}ats, Passeig de Llu\'{\i}s Companys, 23, 08010 Barcelona, Spain}
\affiliation{Institut de F\'{i}sica d’Altes Energies (IFAE), The Barcelona Institute of Science and Technology, Edifici Cn, Campus UAB, 08193, Bellaterra (Barcelona), Spain}

\author[0000-0002-2733-4559]{J.~Moustakas}
\email{jmoustakas@siena.edu}
\affiliation{Department of Physics and Astronomy, Siena University, 515 Loudon Road, Loudonville, NY 12211, USA}

\author{E.~Mueller}
\email{em675@sussex.ac.uk}
\affiliation{Department of Physics and Astronomy, University of Sussex, Brighton BN1 9QH, U.K}

\author[0000-0001-9070-3102]{S.~Nadathur}
\email{seshadri.nadathur@port.ac.uk}
\affiliation{Institute of Cosmology and Gravitation, University of Portsmouth, Dennis Sciama Building, Portsmouth, PO1 3FX, UK}

\author[0000-0001-8684-2222]{J.~ A.~Newman}
\email{janewman@pitt.edu}
\affiliation{Department of Physics \& Astronomy and Pittsburgh Particle Physics, Astrophysics, and Cosmology Center (PITT PACC), University of Pittsburgh, 3941 O'Hara Street, Pittsburgh, PA 15260, USA}

\author[0000-0002-1544-8946]{G.~Niz}
\email{g.niz@ugto.mx}
\affiliation{Departamento de F\'{\i}sica, DCI-Campus Le\'{o}n, Universidad de Guanajuato, Loma del Bosque 103, Le\'{o}n, Guanajuato C.~P.~37150, M\'{e}xico}
\affiliation{Instituto Avanzado de Cosmolog\'{\i}a A.~C., San Marcos 11 - Atenas 202. Magdalena Contreras. Ciudad de M\'{e}xico C.~P.~10720, M\'{e}xico}

\author[0000-0003-3188-784X]{N.~Palanque-Delabrouille}
\email{npalanque-delabrouille@lbl.gov}
\affiliation{IRFU, CEA, Universit\'{e} Paris-Saclay, F-91191 Gif-sur-Yvette, France}
\affiliation{Lawrence Berkeley National Laboratory, 1 Cyclotron Road, Berkeley, CA 94720, USA}

\author[0000-0002-0644-5727]{W.~J.~Percival}
\email{will.percival@uwaterloo.ca}
\affiliation{Department of Physics and Astronomy, University of Waterloo, 200 University Ave W, Waterloo, ON N2L 3G1, Canada}
\affiliation{Perimeter Institute for Theoretical Physics, 31 Caroline St. North, Waterloo, ON N2L 2Y5, Canada}
\affiliation{Waterloo Centre for Astrophysics, University of Waterloo, 200 University Ave W, Waterloo, ON N2L 3G1, Canada}

\author{C.~Poppett}
\email{clpoppett@lbl.gov}
\affiliation{Lawrence Berkeley National Laboratory, 1 Cyclotron Road, Berkeley, CA 94720, USA}
\affiliation{Space Sciences Laboratory, University of California, Berkeley, 7 Gauss Way, Berkeley, CA  94720, USA}
\affiliation{University of California, Berkeley, 110 Sproul Hall \#5800 Berkeley, CA 94720, USA}

\author[0000-0002-2762-2024]{A.~Porredon}
\email{annamaria.porredon@ciemat.es}
\affiliation{CIEMAT, Avenida Complutense 40, E-28040 Madrid, Spain}
\affiliation{Institute for Astronomy, University of Edinburgh, Royal Observatory, Blackford Hill, Edinburgh EH9 3HJ, UK}
\affiliation{Ruhr University Bochum, Faculty of Physics and Astronomy, Astronomical Institute (AIRUB), German Centre for Cosmological Lensing, 44780 Bochum, Germany}
\affiliation{The Ohio State University, Columbus, 43210 OH, USA}

\author[0000-0001-7145-8674]{F.~Prada}
\email{fprada@iaa.es}
\affiliation{Instituto de Astrof\'{i}sica de Andaluc\'{i}a (CSIC), Glorieta de la Astronom\'{i}a, s/n, E-18008 Granada, Spain}

\author[0000-0001-6979-0125]{I.~P\'erez-R\`afols}
\email{ignasi.perez.rafols@upc.edu}
\affiliation{Departament de F\'isica, EEBE, Universitat Polit\`ecnica de Catalunya, c/Eduard Maristany 10, 08930 Barcelona, Spain}

\author{A.~Robertson}
\email{amy.robertson@noirlab.edu}
\affiliation{NSF NOIRLab, 950 N. Cherry Ave., Tucson, AZ 85719, USA}

\author{G.~Rossi}
\email{graziano@sejong.ac.kr}
\affiliation{Department of Physics and Astronomy, Sejong University, 209 Neungdong-ro, Gwangjin-gu, Seoul 05006, Republic of Korea}

\author[0000-0002-0394-0896]{R.~Ruggeri}
\email{rossana.ruggeri@qut.edu.au}
\affiliation{Queensland University of Technology,  School of Chemistry \& Physics, George St, Brisbane 4001, Australia}

\author[0000-0002-9646-8198]{E.~Sanchez}
\email{eusebio.sanchez@ciemat.es}
\affiliation{CIEMAT, Avenida Complutense 40, E-28040 Madrid, Spain}

\author[0000-0002-0408-5633]{C.~Saulder}
\email{csaulder@mpe.mpg.de}
\affiliation{Max Planck Institute for Extraterrestrial Physics, Gie\ss enbachstra\ss e 1, 85748 Garching, Germany}

\author{D.~Schlegel}
\email{djschlegel@lbl.gov}
\affiliation{Lawrence Berkeley National Laboratory, 1 Cyclotron Road, Berkeley, CA 94720, USA}

\author{M.~Schubnell}
\email{schubnel@umich.edu}
\affiliation{Department of Physics, University of Michigan, 450 Church Street, Ann Arbor, MI 48109, USA}
\affiliation{University of Michigan, 500 S. State Street, Ann Arbor, MI 48109, USA}

\author{A.~Semenaite}
\email{asemenaite@swin.edu.au}
\affiliation{Centre for Astrophysics \& Supercomputing, Swinburne University of Technology, P.O. Box 218, Hawthorn, VIC 3122, Australia}

\author[0000-0002-6588-3508]{H.~Seo}
\email{seoh@ohio.edu}
\affiliation{Department of Physics \& Astronomy, Ohio University, 139 University Terrace, Athens, OH 45701, USA}

\author[0000-0002-3461-0320]{J.~Silber}
\email{jhsilber@lbl.gov}
\affiliation{Lawrence Berkeley National Laboratory, 1 Cyclotron Road, Berkeley, CA 94720, USA}

\author{D.~Sprayberry}
\email{david.sprayberry@noirlab.edu}
\affiliation{NSF NOIRLab, 950 N. Cherry Ave., Tucson, AZ 85719, USA}

\author[0000-0002-8246-7792]{Z.~Sun}
\email{szc22@mails.tsinghua.edu.cn}
\affiliation{Department of Astronomy, Tsinghua University, 30 Shuangqing Road, Haidian District, Beijing, China, 100190}

\author[0000-0003-1704-0781]{G.~Tarl\'{e}}
\email{gtarle@umich.edu}
\affiliation{University of Michigan, 500 S. State Street, Ann Arbor, MI 48109, USA}

\author[0000-0003-3841-1836]{M.~Vargas-Maga\~na}
\email{mmaganav@fisica.unam.mx}
\affiliation{Instituto de F\'{\i}sica, Universidad Nacional Aut\'{o}noma de M\'{e}xico,  Circuito de la Investigaci\'{o}n Cient\'{\i}fica, Ciudad Universitaria, Cd. de M\'{e}xico  C.~P.~04510,  M\'{e}xico}

\author{B.~A.~Weaver}
\email{benjamin.weaver@noirlab.edu}
\affiliation{NSF NOIRLab, 950 N. Cherry Ave., Tucson, AZ 85719, USA}

\author[0000-0003-2229-011X]{R.~H.~Wechsler}
\email{rwechsler@stanford.edu}
\affiliation{Kavli Institute for Particle Astrophysics and Cosmology, Stanford University, Menlo Park, CA 94305, USA}
\affiliation{Physics Department, Stanford University, Stanford, CA 93405, USA}
\affiliation{SLAC National Accelerator Laboratory, 2575 Sand Hill Road, Menlo Park, CA 94025, USA}

\author[0000-0002-7305-9578]{P.~Zarrouk}
\email{pauline.zarrouk@lpnhe.in2p3.fr}
\affiliation{Sorbonne Universit\'{e}, CNRS/IN2P3, Laboratoire de Physique Nucl\'{e}aire et de Hautes Energies (LPNHE), FR-75005 Paris, France}

\author[0000-0001-5381-4372]{R.~Zhou}
\email{rongpuzhou@lbl.gov}
\affiliation{Lawrence Berkeley National Laboratory, 1 Cyclotron Road, Berkeley, CA 94720, USA}

\author[0000-0002-6684-3997]{H.~Zou}
\email{zouhu@nao.cas.cn}
\affiliation{National Astronomical Observatories, Chinese Academy of Sciences, A20 Datun Road, Chaoyang District, Beijing, 100101, P.~R.~China}